\documentclass[conference]{IEEEtran}
\IEEEoverridecommandlockouts

\usepackage{amsmath}
\usepackage{graphicx}
\usepackage{hyperref}

\usepackage{adjustbox, multirow}
\usepackage{xcolor}
\usepackage{adjustbox}
\usepackage{booktabs}
\usepackage{algorithm}
\usepackage[noend]{algpseudocode}
\usepackage{url}
\usepackage{subcaption}
\usepackage{todonotes}
\usepackage{stackengine}
\usepackage{epstopdf}

\def\BibTeX{{\rm B\kern-.05em{\sc i\kern-.025em b}\kern-.08em
    T\kern-.1667em\lower.7ex\hbox{E}\kern-.125emX}}

\newif\ifdraft
\ifdraft
    \definecolor{orange}{rgb}{255,0,0}
\else
    \definecolor{orange}{rgb}{0, 0, 0}

\fi

\begin{document}

\title{DynoStore: A wide-area distribution system for the management of data over heterogeneous storage}


\author{\IEEEauthorblockN{Dante D. Sanchez-Gallegos\IEEEauthorrefmark{1}, J. L. Gonzalez-Compean\IEEEauthorrefmark{2}, Maxime Gonthier\IEEEauthorrefmark{3}, Valerie Hayot-Sasson\IEEEauthorrefmark{3}, \\ J. Gregory Pauloski\IEEEauthorrefmark{3}, Haochen Pan\IEEEauthorrefmark{3}, Kyle Chard\IEEEauthorrefmark{3}, Jesus Carretero\IEEEauthorrefmark{1}, and Ian Foster\IEEEauthorrefmark{3}}
\IEEEauthorblockA{\IEEEauthorrefmark{1}Department of Computer Science\\
University Carlos III of Madrid, Leganes, Spain\\
Email: dantsanc@pa.uc3m.es}
\IEEEauthorblockA{\IEEEauthorrefmark{2}Cinvestav Tamaulipas, Cd. Victoria, Mexico}
\IEEEauthorblockA{\IEEEauthorrefmark{3}Department of Computer Science, University of Chicago, Chicago, USA}}


\maketitle

\begin{abstract}
Data distribution across different facilities offers benefits such as enhanced resource utilization, increased resilience through replication, and improved performance by processing data near its source. However, managing such data is challenging due to heterogeneous access protocols, disparate authentication models, and the lack of a unified coordination framework. This paper presents DynoStore, a system that manages data across heterogeneous storage systems. At the core of DynoStore are data containers, an abstraction that provides standardized interfaces for seamless data management, irrespective of the underlying storage systems. Multiple data container connections create a cohesive wide-area storage network, ensuring resilience using erasure coding policies. Furthermore, a load-balancing algorithm ensures equitable and efficient utilization of storage resources. We evaluate DynoStore using benchmarks and real-world case studies, including the management of medical and satellite data across geographically distributed environments. Our results demonstrate a 10\% performance improvement compared to centralized cloud-hosted systems while maintaining competitive performance with state-of-the-art solutions such as Redis and IPFS. DynoStore also exhibits superior fault tolerance, withstanding more failures than traditional systems.
\end{abstract}


\section{Introduction}




Modern scientific applications generate vast amounts of data from highly distributed data sources such as sensor networks~\cite{boubiche2018big}, scientific instruments~\cite{stephens2015big}, and medical devices~\cite{dash2019big}. 
Efficiently managing this data is critical for enabling real-time insights, scientific collaboration, and robust decision-making. While traditional storage systems efficiently manage data within a single location, they struggle to meet the demands of distributed environments due to finite capacity and susceptibility to failures~\cite{ahmadian2018investigating}. 

Distributed storage systems mitigate these limitations by spreading data across multiple locations, ensuring high availability and scalability. Many applications now require such multi-storage setups, such as medical diagnostics and research~\cite{draganski2019nation}, scientific collaboration \cite{kahn2011future}, and earth observation~\cite{kotsev2020spatial}. These systems face challenges such as protocol heterogeneity, inconsistent authentication, and lack of unified coordination. While public cloud solutions like AWS Storage Gateway~\cite{mishra2023advanced} address some of these issues, they introduce concerns such as vendor lock-in~\cite{opara2016critical}, high operational costs, and limited compatibility with edge infrastructure~\cite{liu2017minimum}.

Thus, there is a need for solutions that enable scientists and organizations to interconnect multiple storage systems \cite{suter2023driving}. These solutions must:  \textit{i)} enable the transparent management of various storage systems (e.g., Ceph~\cite{aghayev2019file}, Lustre~\cite{madireddy2018machine}, HDFS~\cite{ghazi2015hadoop}, or S3~\cite{bornholt2021using}) as a single distributed system; \textit{ii)} efficiently manage the available storage resources; \textit{iii)} enable flexible access to data in scenarios where the location of data may change over time; \textit{iv)} provide access to data even in scenarios where individual storage locations fail \cite{zhang2020exploring}; and \textit{v)}  ensure data security to avoid unauthorized access.

Considering the above, this paper presents DynoStore, simplifying data management across heterogeneous storage systems. DynoStore uses management units called \textit{data containers} as a foundational abstraction, providing standardized interfaces to interconnect storage systems seamlessly.  Data containers implement an object store-like interface seamlessly deployed on an underlying storage system. 
DynoStore connects multiple data containers, creating a cohesive wide-area storage network that can also be described as a geographically distributed object store system. DynoStore enables efficient resource utilization through load balancing and ensures resilience with erasure coding-based policies.


We implemented DynoStore following a modular microservice architecture, which multiple clients and systems can consume.  We evaluate DynoStore through various benchmarks to measure its efficiency and scalability, and conduct two case studies for the processing and storing of medical and satellite imagery, creating a secure and reliable distribution network to enable collaboration between organizations and scientists. 

In summary, the main contributions of this paper are:

\begin{itemize}
    \item DynoStore, a system to build wide-area distribution networks combining heterogeneous and distributed data storage systems.
    \item A data management framework incorporating resilience and load-balancing policies to maximize fault tolerance and efficiency.
    \item An experimental evaluation demonstrating DynoStore's performance, including a 10\% improvement over centralized solutions and superior fault tolerance compared to state-of-the-art systems.
\end{itemize}

The rest of this paper is as follows: Section \ref{sec:relatedwork} reviews related work. Section \ref{sec:design} describes the design principles of DynoStore. Section \ref{sec:eva} presents a performance evaluation of DynoStore. Section \ref{sec:disc} discusses the main results of DynoStore.
Finally, Section \ref{sec:conc} summarizes our contributions.

\color{orange}
\section{Background and Related Work}
\label{sec:relatedwork}

\textbf{Distributed filesystems} like Lustre \cite{madireddy2018machine} and HDFS \cite{ghazi2015hadoop}  have been widely adopted for managing large workloads within an organization. Ad-hoc systems like Expand \cite{garcia2003design} offer customizable storage solutions, providing flexibility and fault-tolerance \cite{munoz2024fault} to meet specific application needs. These systems ensure scalability and high availability through replication, making them well-suited for batch processing. However, their reliance on cluster-based architectures limits their effectiveness in real-time and geographically distributed scenarios. For instance, Lustre and HDFS do not natively support heterogeneous storage backends. DynoStore addresses these gaps by introducing data containers that enable seamless interconnection of diverse storage systems while ensuring data reliability and distributed access.

\textbf{Object stores} like DAOS \cite{hennecke2020daos}, Ceph \cite{aghayev2019file}, and S3 \cite{bornholt2021using} offer a more flexible model by decoupling data from metadata and enabling storage across multiple regions. While Ceph and DAOS offer interfaces for connecting heterogeneous nodes, all parties must meet different requirements. For example, Ceph requires that all parties adopt the same file system, whereas DAOS requires NVM hardware.  Public cloud-based solutions like S3~\cite{bornholt2021using}, on the other hand, impose vendor lock-in and confidentiality risks, limiting their applicability for critical applications. DynoStore avoids these limitations through its infrastructure-agnostic design and the use of standardized data containers, which can be added or removed without complex configuration.

\textbf{Content-delivery networks} (CDNs), including Amazon CloudFront \cite{cloudfront}, Azure CDN \cite{azure}, and Fastly \cite{fastly} as well as decentralized systems like IPFS (InterPlanetary File System) \cite{benet2014ipfs} efficiently manage data replication and distribution. However, IPFS lack robust fault-tolerance mechanisms. For example, IPFS relies on a peer-to-peer model, making data unavailable if a storing peer fails. DynoStore fills this gap by incorporating erasure coding for resilience and load-balancing policies to ensure fair utilization of storage resources across distributed environments.

\textbf{Unified Data Management Platforms} like 
Rucio \cite{barisits2019rucio} use a metadata catalog to organize and locate data across multiple storage systems. Rucio handles the placement and searching of data through these systems. While Rucio supports the definition of data replication rules, users have to define these rules without considering the utilization and capacity of resources. Similarly, iRODs \cite{irods} connects multiple distributed storage environments---implementing replication and load-balancing to guarantee data availability.  Data replication produces high storage overhead, which strains resources with limited storage availability. DynoStore instead implements erasure coding and data placement strategies based on utilization to create a fair data distribution. In this sense, DynoStore is similar to OceanStore~\cite{kubiatowicz2000oceanstore}, but adds new features, like managing data across existing storage systems and infrastructure, without requiring specialized infrastructure.
\color{black}

\section{DynoStore: design principles}
\label{sec:design}

In this section, we describe the design principles of DynoStore. 
We designed DynoStore to meet the following requirements: \textit{i)} manage the storage of data in different locations with different characteristics in terms of filesystem, storage capacity, and reliability; 
\textit{ii)}  maintain efficient usage of the underlying storage resources by load-balancing data; 
\textit{iii)} keep access to data even across multiple storage silos, creating a single storage system or data pool; 
\textit{iv)} create a resilient and secure system to keep access to data even where individual storage locations fail;
and \textit{v)} process data in multiple locations and connect them through a single data system.

\subsection{Data containers}

DynoStore introduces data containers as the foundational abstraction for managing data across distributed and heterogeneous storage systems. A data container serves as a middleware layer, providing standardized interfaces that abstract the complexities of the underlying storage infrastructure. This enables seamless integration with diverse storage backends, such as Ceph, HDFS, and NFS.

Each data container includes the following components: \textit{i)} \textbf{Access Interfaces}:  REST APIs for data operations, such as upload, download, delete, and search. Requests are authenticated using OAuth tokens to ensure secure access; \textit{ii)} \textbf{Monitor}: A service that checks the state of the underlying storage system; \textit{iii)} \textbf{Caching Layer}: Implements a Least Recently Used (LRU) caching policy to minimize access latency and reduce interactions with the underlying storage system. \textcolor{orange}{When a new object arrives, it is written into memory and the local storage system. This avoids losing data if the storage container fails. Objects exceeding the available memory size are written directly to the filesystem. This caching layer is complementary to the caching natively implemented by storage systems. While the storage system improves internal access to data, the data container's caching layer reduces the number of interactions with the storage system.}

\textcolor{orange}{Administrators deploy data containers by installing the DynoStore agent and providing a configuration file that specifies the container's name, storage path, and access parameters. For instance, a data container on HDFS might require a file path and backend details, while one on NFS only needs a directory path. This plug-and-play model simplifies deployment and troubleshooting, enabling organizations to dynamically add or remove containers as needed.}

\subsection{Management services}


\begin{figure}[t]
    \centering
    \includegraphics[width=0.8\linewidth]{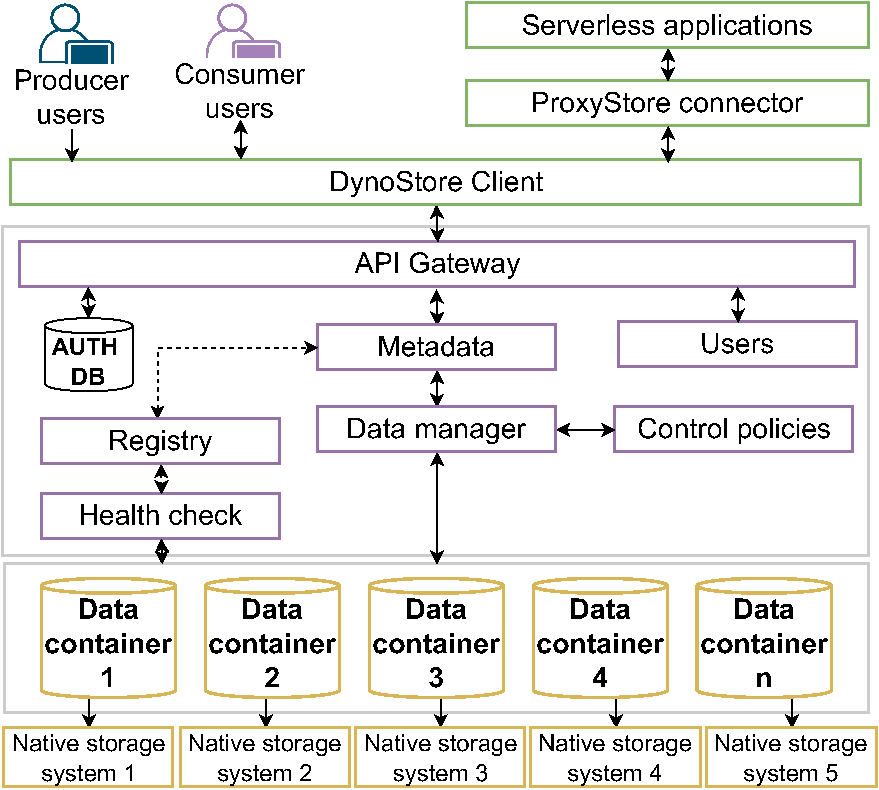}
    \caption{DynoStore architecture. 
    }
    \label{fig:archi}
    \vspace{-1.5em}
\end{figure}

DynoStore's management services form the backbone of its distributed architecture, enabling coordination, scalability, and secure data operations across multiple storage endpoints using data containers. \textcolor{orange}{Management services simplify coordination across distributed storage endpoints, ensuring consistent metadata, efficient resource utilization, and streamlined enforcement of control policies across a heterogeneous environment.} These services are implemented as a modular microservices architecture, as shown in Figure \ref{fig:archi}, and include components for authentication, metadata management, container registration, health monitoring, and policy enforcement. 
The gateway service acts as the entry point for client requests. It validates user credentials through OAuth tokens and routes authorized requests to the appropriate backend services. By efficiently handling concurrent requests, the gateway ensures seamless interactions between users and DynoStore.

The metadata service maintains detailed records of all objects in the system, including their UUIDs, locations, sizes, and ownership. This information enforces namespace structures, manages permissions, and maintains consistency during updates. For example, when a user uploads an object, the metadata service records its UUID and storage location, ensuring efficient lookup during future operations.
The registry tracks all active data containers in the system. Administrators can dynamically add or remove containers, and the registry updates its records in real-time to reflect these changes. This dynamic tracking ensures that new storage resources can be seamlessly integrated into the system.
The health check service continuously monitors the availability and performance of all data containers. When a container becomes unavailable, the service dynamically reallocates operations to healthy containers, maintaining system reliability and resilience.

DynoStore enforces control policies for data resilience and load balancing. These policies leverage utilization metrics and erasure coding to create a fair utilization of resources while ensuring data remains accessible, even during infrastructure failures.

\color{orange}
\subsection{Services scalability and fault tolerance}

Organizations deploy on infrastructure accessible by data container---for example, in cloud instances with a public IP address. They can scale each service in DynoStore across multiple nodes.  Scale-in is implemented using a multi-threading approach, enabling efficient handling of concurrent requests. Scale-out is performed by deploying multiple replicas of a service across distributed machines. DynoStore employs Paxos \cite{howard2020paxos} to coordinate operations among service replicas to support scalability and fault tolerance (see Section \ref{sec:consistency}). 

\color{black}

\section{Data management model}

\textcolor{orange}{DynoStore’s data management model is designed to address the challenges of securely and efficiently handling data across heterogeneous and distributed storage systems. The model introduces a logical namespace structure for organizing data, enforces strong consistency through Paxos-based replication, and ensures resilience with erasure coding. Additionally, a load-balancing algorithm manages data placement across containers, while robust authentication and integrity checks maintain security. This section describes these features in detail, focusing on namespaces, consistency, placement, and resilience policies.}

\subsection{Data namespace and permissions}

\begin{figure}[t]
    \centering
    \includegraphics[width=0.75\linewidth]{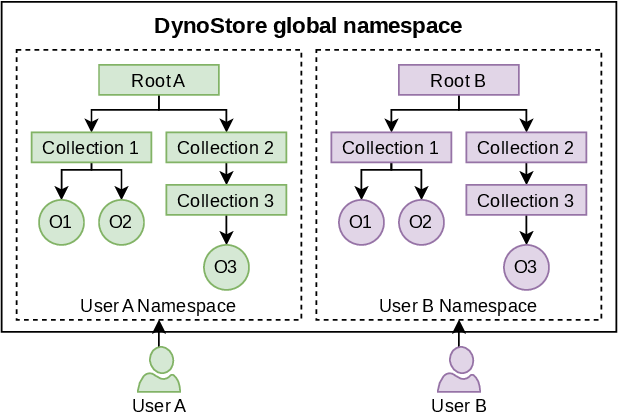}
    \caption{DynoStore's data namespace structure.}
    \label{fig:treeorganization}
    \vspace{-1.5em}
\end{figure}

\textcolor{orange}{As can be observed in Figure \ref{fig:treeorganization}, DynoStore organizes data into virtual namespaces, which provide isolated environments for each user. A namespace contains all objects uploaded by the user, such as medical images or satellite data, along with their associated metadata." Within a namespace, users can create collections, which act as hierarchical groupings for organizing related objects. For instance, a user managing satellite images might create collections for specific regions or timeframes.}

Objects within collections are uniquely identified by a user-defined name and a UUID generated by DynoStore’s metadata service to ensure global uniqueness and traceability. By default, all objects in a namespace are stored in a root collection named after the user (e.g., \texttt{/UserA}). Users can create nested collections by specifying the name or UUID of an existing collection during creation. Users must provide the collection's absolute path to upload an object to a nested collection, similar to a Unix directory structure.
For example, a collection of satellite imagery from a specific region might be organized as \texttt{/UserA/Satellite/Region1/Scene2}.

DynoStore also enforces permissions at both the object and collection levels. Permissions are inherited by default, meaning access granted to a collection applies to all its subcollections and objects unless overridden. This simplifies access control for complex hierarchies. For instance, granting read access to \texttt{/UserA/Collection1} automatically extends to \texttt{/UserA/Collection1/Subcollection2} and its objects.
{The combination of namespaces, collections, and hierarchical paths gives users a flexible and secure framework for organizing their data.

\subsection{Data update and consistency model} \label{sec:consistency}

\textcolor{orange}{To ensure data integrity and simplify consistency management, objects within DynoStore namespaces are immutable. Once uploaded, an object cannot be modified directly. Instead, users can update objects by uploading a new version, which is assigned a new UUID by DynoStore’s metadata service. The metadata is then updated to reference the latest version. This versioning system enables users to roll back to earlier versions if needed, providing both flexibility and reliability.}

\textcolor{orange}{DynoStore also implements a garbage collection mechanism to manage outdated object versions. By default, older versions are retained for 30 days before being automatically deleted. Users can customize this retention period based on their requirements. For example, a user storing medical data may retain older versions to comply with regulatory standards.}

\textcolor{orange}{To maintain consistency, DynoStore ensures strong read-after-write consistency. When an object is updated, read operations are temporarily locked until the metadata is fully updated. Consistency is managed using the Paxos consensus algorithm in scenarios where the metadata service is replicated. This process ensures that all replicas agree on the state of the metadata, even in the presence of failures. The Paxos-based data update process includes the following steps: \textit{i)} A client sends an update request to a metadata replica (proposer); \textit{ii)} The proposer sends a message containing the current UUID of the object and a timestamp of the request to the other replicas; \textit{iii)} Each replica checks the timestamp and if the timestamp is greater than the last recorded update for the object, the replica responds with an acceptance message; \textit{iv)} After receiving acceptance messages from a majority of replicas, the proposer updates the object and broadcasts the new UUID and timestamp to all replicas. }

\textcolor{orange}{This approach ensures strong consistency, even during partial failures, by coordinating updates across replicas. For instance, consider a satellite imagery dataset where an updated image is uploaded to replace an older version. DynoStore ensures that subsequent reads always access the latest version, even if some replicas experience failures.}

\textcolor{orange}{By combining immutability, versioning, garbage collection, and strong consistency, DynoStore provides a reliable and efficient framework for managing updates in distributed environments.}

\subsection{Data placement and load-balancing} \label{sec:loadbalancing}

In DynoStore, data placement is determined by a load-balancing algorithm based on a metric called utilization factor (UF) \cite{carrizales2024structmesh}. This algorithm aims to efficiently use the storage resources, producing a fair distribution for each data container while avoid  overloading individual containers.

The utilization factor measures the available space in a storage container with respect to the total storage capacity of all containers in the system. Given an object $o$, we first calculate the UF of the memory ($U(x)_{mem}$) and storage ($U(x)_{fs}$) resources available for each data container as follows:
\begin{equation}
\begin{split}
    U(x)_{mem} &= 1 - \frac{M(x)_{total} - [M(x)_{available} - |o|]}{M(x)_{total}}, \\
    U(x)_{fs} &= 1 - \frac{S(x)_{total} - [S(x)_{available} - |o|]}{S(x)_{total}},
\end{split}
\end{equation}
where $M(x)_{available}$ and $S(x)_{available}$ are the available memory and storage in container $x$; $M(x)_{total}$ and $S(x)_{total}$ are the total memory and storage capacities; and $|o|$ is the size of the object to store.

The container with the lowest combined utilization factor is selected:
\begin{equation}\label{eq:choosing}
    s = \min_{x \in \mathcal{D}} \left( w_1U(x)_{mem} + w_2U(x)_{fs} \right),
\end{equation}
\textcolor{orange}{where $s$ is the selected data container, $x$ is a data container, $\mathcal{D}$ is the set of available data containers, and $w_1$ and $w_2$ are adjustable weights that prioritize memory or storage usage based on application requirements. For example, in a medical imaging scenario where data must be preserved long-term, administrators may assign a higher weight to $w_2$ to favor containers with more available storage. In contrast, a higher $w_1$  weight might prioritize containers with available memory for short-term data caching.}

\textcolor{orange}{DynoStore’s load-balancing algorithm is extensible, allowing additional metrics like bandwidth, latency, or cost to be integrated. This flexibility ensures that the system can adapt to diverse workloads and deployment environments, optimizing performance and resource utilization.}

\subsection{Data resilience} \label{sec:resilience}

To ensure fault tolerance, DynoStore implements a data resilience policy based on an erasure coding technique called \textit{information dispersal algorithm~\cite{rabin1990information}}. This approach divides an object into $n$ chunks, including $k$ data chunks and $n-k$ parity chunks. The object can be reconstructed using any $k$ chunks, allowing the system to tolerate up to $n-k$ failures.

\begin{algorithm}[t]
\caption{Data encoding process.}\label{alg:encode}
\begin{algorithmic}[1]
\scriptsize
\Function{encode}{$o$, $n$, $k$}
  \State $\mathcal{D}$ = \Call{GetAvailableDC}{$n$} \label{ln:datacontainers}
  \If{$|\mathcal{D}| < n$}
    \State
    \Return \Call{Error}{Not enough containers available.} 
  \Else
    \State $\mathcal{C}$ = \Call{split}{$o$, $n$, $k$} \label{ln:split}
    \State $h_{o}$ = \Call{sha256}{$o$} \label{ln:sha}
    \For{$i=0;i<n;i++$}
        \State $p$ = \Call{Pack}{$h_{o}$, $\mathcal{C}[i]$} \label{ln:pack}
        \State \Call{Upload}{$p$, $\mathcal{D}[i]$} \label{ln:upload}
    \EndFor
  \EndIf
\EndFunction

\end{algorithmic}

\end{algorithm}

Algorithm \ref{alg:encode} presents the process of storing an object under this data resilience policy. Line \ref{ln:datacontainers} shows that the first step is to retrieve $n$ data containers using the load-balancing strategy described in Section \ref{sec:loadbalancing}. An error is produced if there are insufficient data containers in $\mathcal{D}$. Then, in Line \ref{ln:split}, the input object $o$ is split into $n$ chunks $\mathcal{C}$ and in Line \ref{ln:sha}, the hash $h_o$ of the original object is calculated using a SHA3-256 functions. This hash is used during decoding to verify the object's integrity and ensure it has not been modified during transportation and storage. The hash is packed with each chunk $\mathcal{C}[i]$ (Line \ref{ln:pack}), and then these packages are uploaded to the data containers $\mathcal{D}$ (Line \ref{ln:upload}).

\begin{algorithm}[t]
\caption{Data decoding process to retrieve an object.}\label{alg:decode}
\begin{algorithmic}[1]
\scriptsize
\Function{decode}{$id$}
    \State $k$ = \Call{GetKFromMetadata}{$id$} \label{ln:chunks}
    \State $\mathcal{C}$ = \Call{RetrieveChunks}{$id$, $k$}  \label{ln:download}
    \If{$|\mathcal{C}| \geq k$} \label{ln:verify}
        \State $ro$ = \Call{merge}{$\mathcal{C}$, $k$} \label{ln:merge}
        \State $h_o$ = \Call{readObjHash}{$\mathcal{C}[0]$}
        \State $h_{ro}$ = \Call{sha256}{$ro$}
        \If{$h_o == h_{ro}$} \label{ln:comp}
            \State
            \Return \Call{Error}{} 
            \Comment{The hashes are different.}
        \EndIf
        \State
        \Return $ro$
    \Else
        \State
        \Return \Call{Error}{} 
        \Comment{Not enough chunks.}
    \EndIf
\EndFunction

\end{algorithmic}
\end{algorithm}

Algorithm \ref{alg:decode} presents the process of downloading a set of chunks and decoding them to retrieve the original object $ro$. First, in Line \ref{ln:chunks}, the number of chunks $k$ used to code the data is obtained, and in Line \ref{ln:download}, the $k$ chunks are retrieved by passing the identifier $id$ of an object stored in DynoStore as a reference. If enough chunks $\mathcal{C}$ are downloaded (Line \ref{ln:verify}), they are merged into a single object $ro$ (Line \ref{ln:merge}). Finally, in Line \ref{ln:comp}, the reconstructed object $ro$ hash is compared with the hash calculated during the encoding process. If they are different, an error is produced. Otherwise, $ro$ is returned.

\textcolor{orange}{The resilience policy’s fault tolerance depends on the parameters $n$ (total chunks) and $k$ (chunks needed for recovery). For example, $n=10,k=7$ tolerates up to 3 failures, whereas $n=12,k=8$ tolerates up to 4 failures. By distributing chunks across containers in different geographic locations, DynoStore enhances fault tolerance against localized failures, ensuring data availability even in adverse conditions. }

\subsection{Data Security}

DynoStore employs a multi-layered approach to security that considers access control, integrity, and confidentiality.

\subsubsection{Access control}

DynoStore uses an OAuth authentication model to validate the requests of both clients and data containers. When a user initiates a request, the authentication service issues an OAuth token, encapsulating user credentials and permissions. This token is validated by the API gateway for every request, ensuring that only authorized users can access or modify data. 

\subsubsection{Data Integrity and Confidentiality}

To ensure data integrity, DynoStore computes SHA3-256 hashes of all objects during upload and stores these hashes in the metadata service. When an object is retrieved, the system re-computes its hash and compares it to the stored value, detecting any corruption or tampering. 

Furthermore, the resiliency policy implemented in DynoStore, combined with the load-balancing algorithm, guarantees that each data container only contains a chunk of the data, ensuring that unauthorized users cannot access objects.

Point-to-point confidentiality can be enabled in DynoStore's client to encrypt objects before an upload operation. DynoStore's client implements an AES-256 encryption to safeguard sensitive objects (e.g., medical data) during transport.

\section{Implementation details}

DynoStore is implemented as a microservice architecture in Python 3. This modular approach allows users to deploy the different architecture components through distributed machines in a cluster and scale them independently. These microservices are encapsulated in Docker virtual containers. Users can also directly deploy DynoStore from its source code. DynoStore access interfaces are developed as REST APIs; thus, data uploading and downloading are implemented using HTTP. In this first version, we opted for this protocol to reduce the complexity of its use on the client, as it is widely allowed across firewalls and NATs. Nevertheless, we are exploring the integration of DynoStore with faster data transfer tools like IPFS, and Globus Transfer \cite{chard2017globus}. 

Access to DynoStore is performed through a client that connects to the gateway to push and pull data.  The DynoStore client implements functions to perform basic data operations such as push, pull, exists, and evict. We implemented this client using Python, which can be used as a command-line program or as a library that can be integrated into applications to perform push and pull operations directly. Furthermore, we implemented a ProxyStore connector \cite{pauloski2023accelerating} to simplify integration in distributed task-based applications. 
ProxyStore is a Python framework that enables the transparent management of Python objects as a Proxy, which, in the case of DynoStore, is a reference to an object stored in a data container.  Thus, a Python program can consume and process this reference as a native Python object, but it is stored in a remote location. This integration enables using DynoStore as storage for various task-based applications, such as workflows and FaaS applications. 

\section{Evaluation}
\label{sec:eva}

In this section, we present the evaluation of DynoStore in three phases.
\textit{i)} We perform microbenchmarks to assess the performance and efficiency of our solution in the general case.
\textit{ii)} We focus on a case study for managing lung tomography images. 
\textit{iii)} We test the robustness of DynoStore with a use case focused on managing satellite imagery across the computing continuum. 



\subsection{Dataset}

We used three datasets to conduct our experimental evaluation. The first dataset used in the microbenchmarks consists of synthetic objects with random content, ranging in size from 1 MB to 10,000 MB. The second dataset contains 119,288 breast and lung tomography images, totaling 21 GB. The lung tomography images, acquired from the publicly available LCTSC dataset \cite{lctsc}, account for 10 GB, while the breast tomography images, from the QIN-Breast dataset \cite{li2016data}, make up 11 GB. The third dataset consists of 4,852 satellite images (MODIS and LandSat), totaling 1.2 TB.

\subsection{Infrastructure} \label{sec:infraestructure}

\begin{table}[t]
\centering
\caption{Characteristics of the infrastructure.}
\label{tab:inf}
\resizebox{\columnwidth}{!}{%
\begin{tabular}{ccccc}
\toprule
\textbf{Server} & \textbf{Location} & \textbf{Memory (GB)} & \textbf{Storage (TB)} & \textbf{\# CPU} \\ \midrule
Client1 & Madrid, Spain & 125  & 0.4  & 80 \\
DSEndpoints1-10  & Chameleon  & 251  & 1  & 96 \\
DSEndpoints11-15  & AWS (North-Virginia) & 0.4  & 80 & 8 \\
DSEndpoints16-20  & AWS (North-Virginia) & 0.4  & 80 & 8 \\
Metadata  & Chameleon & 16 & 1 & 8 \\ 
GCEndpoint1 & Chameleon  & 251  & 1  & 96 \\
GCEndpoint1 & Victoria, Mexico & 125 & 0.2 & 48 \\
\bottomrule  
\end{tabular}
}
\end{table}

To show the feasibility of DynoStore in managing data across distributed and heterogeneous environments, we deploy data containers across different endpoints. The characteristics of these endpoints are shown in Table \ref{tab:inf}. DSEndpoints1-10 are a set of storage nodes in the Chameleon cloud \cite{keahey2020lessons}. 
Half of these nodes are in the CHI@TACC region, and the rest are in the CHI@UC. DSEndpoints11-15 are AWS-EC2 virtual machines with Elastic Block Store (EBS) using solid-state drives (SSD). These nodes are also connected to a filesystem implemented using Amazon FSX for Lustre with a throughput of 300 MB/s and a total size of 1.2 TB. Similarly, DSEndpoints16-20 are also AWS-EC2 virtual machines with EBS using traditional hard disk drives (HDD). Metadata is a machine that deploys DynoStore's backend services (gateway, metadata, and access control). GCEndpoint1 and GCEndpoint2 are Globus Compute Endpoints in Chameleon and a private cluster, respectively. 

\subsection{Performance evaluation under different microbenchmarks}


Here, we evaluate DynoStore's performance with different microbenchmarks by measuring the time required to upload/download data with DynoStore, the deployment time of data containers, and the performance of the resilience policy.  


\subsubsection{Data container deployment}

\begin{figure}
    \centering
    \includegraphics[width=0.65\linewidth]{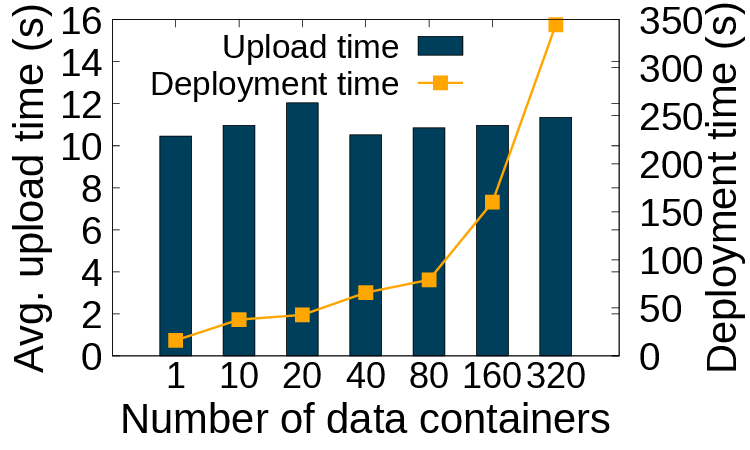}
    \caption{Time to deploy a varying number of data containers on the Chameleon Cloud and average time per request observed to upload 100 objects of 100 MB.}
    \label{fig:deployconts}
    \vspace{-1.5em}
\end{figure}

We first evaluated the time required to deploy a varying number of data containers. 
Figure \ref{fig:deployconts} illustrates the relationship between deployment time and the number of data containers, with the right y-axis representing the total deployment time on ten bare-metal instances on Chameleon. The same number of data containers were deployed for each instance, emulating a scenario where ten geographically distributed organizations manage these containers. Figure \ref{fig:deployconts} also shows on the left y-axis the average time to upload 100 objects of 100 MB to the system. As expected, the deployment time increases as more containers are in the system. At the same time, the time to upload the data remains almost constant for each configuration. Thus, the number of containers does not significantly impact the system's performance, as DynoStore's load-balancer distributes the input requests to the different data containers. 

\subsubsection{Comparing data resilience policies}
\begin{figure}
    \centering
    \includegraphics[width=0.65\linewidth]{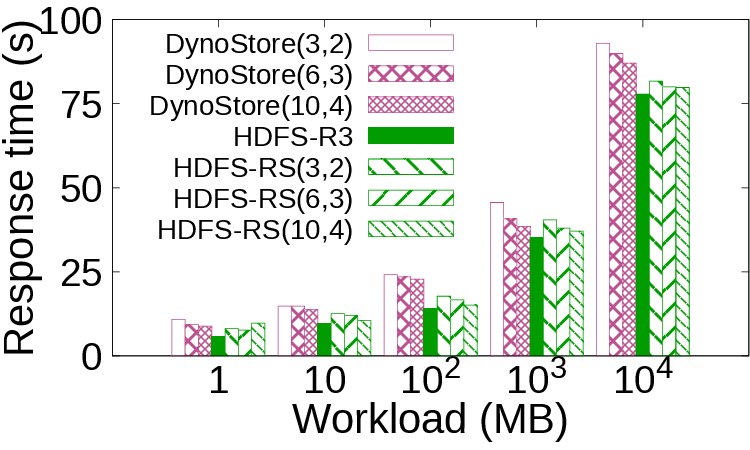}
    \caption{Response time when uploading different data sizes varying resilience configurations in DynoStore and HDFS.}
    \label{fig:resilience}
    \vspace{-1.5em}
\end{figure}


In the next experiment, we evaluated and compared the resilience policy implemented in DynoStore and HDFS. This last one uses both replication and Reed-Solomon (RS) erasure code. 
For HDFS, we used RS(3,2), RS(6,3), and RS(10,4) policies, which support two, three, and four failures, respectively.  Moreover, in HDFS, we evaluated the three-copy replication strategy that supports two failures. Meanwhile, in DynoStore, we evaluated configurations of $n=\{10, 6, 3\}$ and $k=\{4, 3, 2\}$ supporting the same number of failures as HDFS. 


Figure~\ref{fig:resilience} shows, on the y-axis, the response time observed when downloading different data sizes using DynoStore and HDFS with these configurations. In general, HDFS-R3, the one using replication, is the fastest configuration because replication involves fewer computations than erasure coding. Nevertheless, comparing HDFS-RS and DynoStore, we observed competitive response times due to the similar number of operations: data upload + chunking + parity blocks calculation + storage. Note that HDFS and DynoStore scopes are different, as the first one is developed for efficient local storage in a cluster. In contrast, DynoStore manages data storage across various distributed storage locations.  

\subsubsection{Measuring data uploading/downloading costs}

We measure the time required to upload and download data to/from a system created with DynoStore. We consider two scenarios: \textit{i)} the ``\textit{Regular}'' scenario, which stores the data without any resilience policy. Thus, each object is stored on a single server without replicating it nor chunking it. This configuration is our baseline for measuring the overhead of the resiliency policy implemented in DynoStore. \textit{ii)} ``Resilience'' is the second scenario with a configuration of $n=10, k=7$, which supports up to three failures.  We considered two environments for this evaluation: The first emulates a client geographically near the storage system. We refer to this environment as Chameleon $\to$ Chameleon. In the second node, clients are deployed in a remote geographic location on a private cluster in Madrid. This second environment is denoted as Madrid $\to$ Chameleon. 

\begin{figure}[t]
    \centering
         \includegraphics[width=0.7\linewidth]{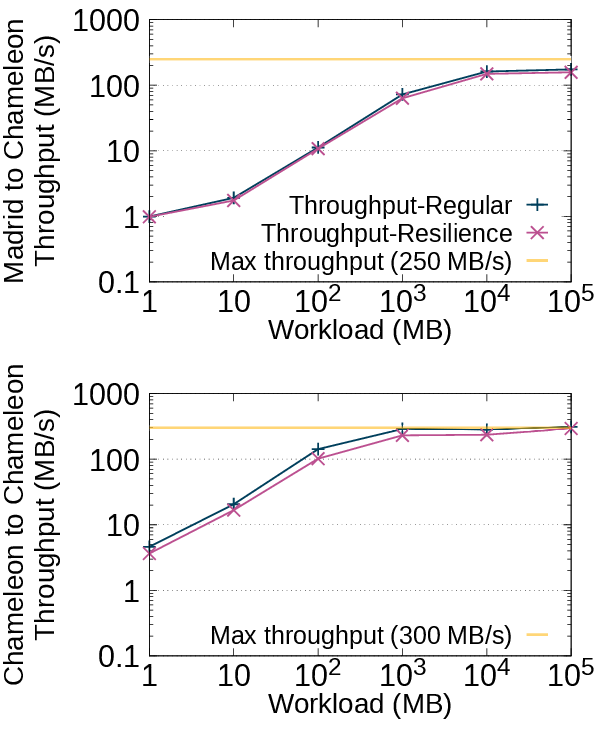}
    \caption{Throughput measurement for uploads of different workload sizes with and without resilience.  
    }
    \label{fig:times}
    \vspace{-1.5em}
\end{figure}

Figure~\ref{fig:times} shows, on the y-axis, the throughput to upload different workloads (x-axis) using Regular and Resilience configurations.  For each workload size in the figure, we sent 100 requests to determine the average throughput. The max throughput was measured using \texttt{iperf}.  We can observe that the Resilience configuration generally produces a lower throughput than the Regular configuration. For example, in the Madrid $\to$ Chameleon, DynoStore uploads 1000 MB of data in 8.9 seconds under the Regular configuration. Meanwhile, with the Resilience configuration, it took 9.2 seconds. This represents a difference of 17\% in the response time. Similarly, during the data download, we can observe that the Regular configuration yields a response time of 9.4 seconds, whereas the Resilience configuration yields 10.5 seconds. This increase in time is expected, as the Resilience configuration is performing additional tasks on the server side to \textit{i)} split the objects into $n$ chunks, \textit{ii)} add redundancy, and \textit{iii)} upload the $n$ chunks to $n$ different data containers, which involves handling more connections than uploading only a single object without chunking.  Thus, this difference in the time is the overhead added by the resilience policy implemented in DynoStore.

\begin{figure}[t]
    \centering
         \includegraphics[width=0.7\linewidth]{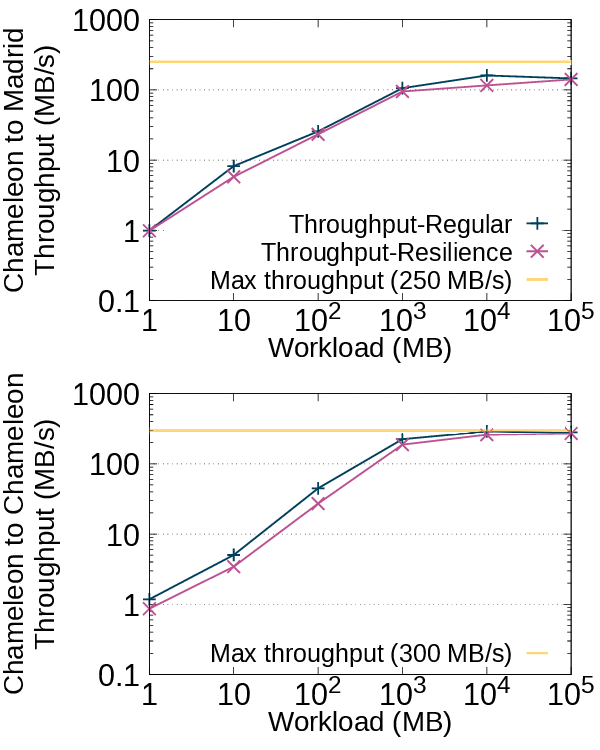}
    \caption{Throughput measured when downloading different volumes of data through different locations.}
    \label{fig:timesdown}
\end{figure}

\subsubsection{Improving data operations using parallel channels}

\begin{figure}[t]
    \centering
    \begin{subfigure}[b]{0.49\textwidth}
    \centering
         \includegraphics[width=0.65\linewidth]{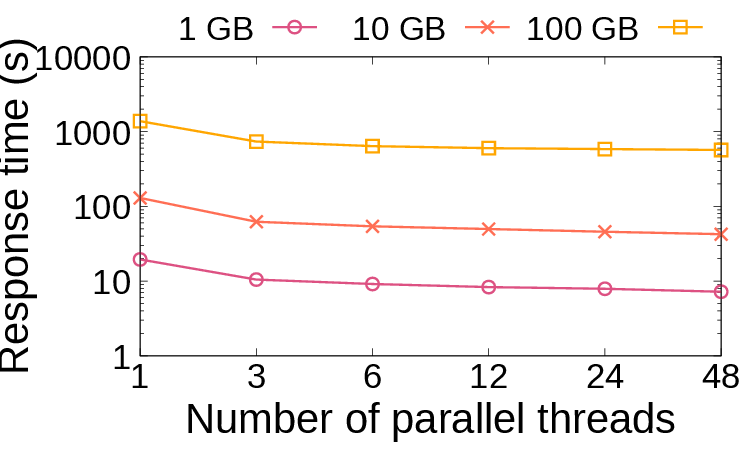}
        \caption{Upload.}
        \label{fig:concup}
     \end{subfigure}
     \hfill
     \begin{subfigure}[b]{0.49\textwidth}
         \centering
         \includegraphics[width=0.65\textwidth]{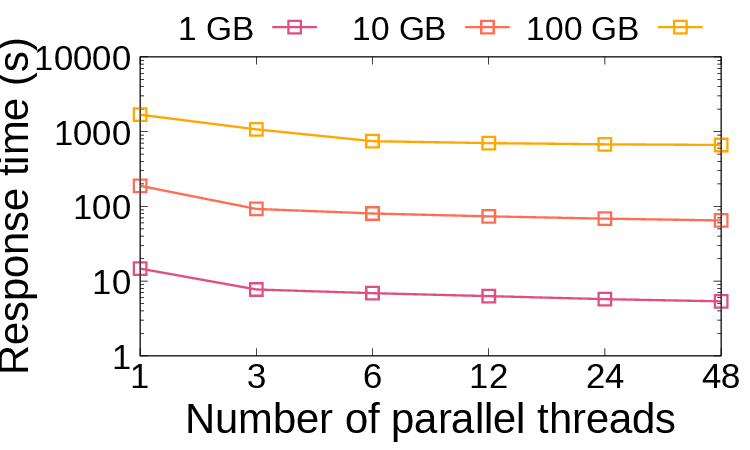}
         \caption{Download.}
         \label{fig:concdown}
     \end{subfigure}
    \caption{Response time of data operations with various threads.}
    \label{fig:parallel}
\end{figure}

In the next experiment, we evaluated the performance of a parallel data upload and download scheme in DynoStore using the Madrid $\to$ Chameleon environment. Figure \ref{fig:parallel} presents the response time (y-axis) for uploading and downloading 100 objects, each larger than 1 GB, as the number of parallel threads increases (x-axis). The number of threads represents the number of channels concurrently opened for data transfer between the client and DynoStore's storage system. On the server side, each channel is handled by a separate replica instance of DynoStore's management services.
We observe a reduction of 58\% when uploading 100 GB of data with 48 threads instead of one. This enables DynoStore to handle large workloads using the resources available on both the client and server sides, accelerating the transfer of data.

\subsubsection{Deployment in a public cloud environment}
\begin{figure}[t]
    \centering
         \includegraphics[width=.65\linewidth]{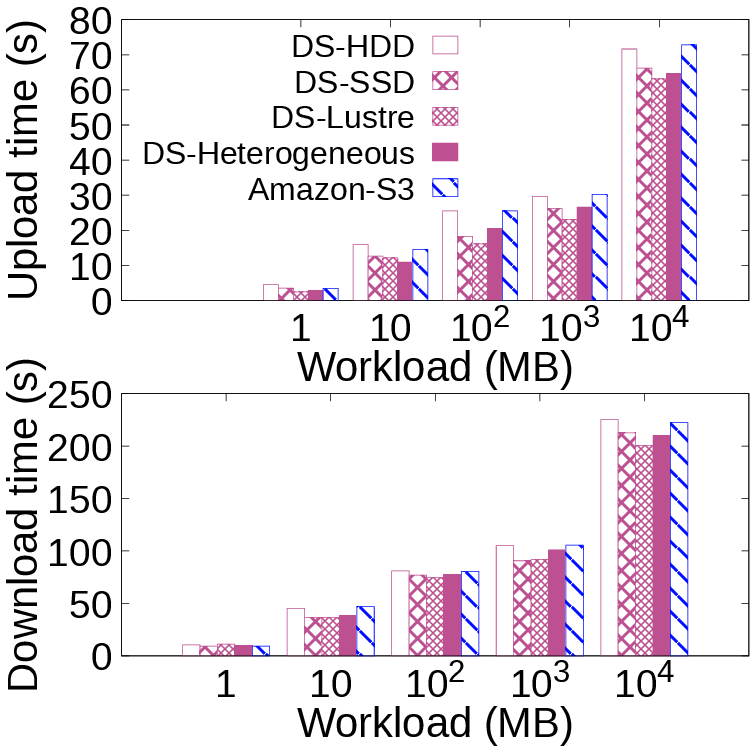}
    \caption{Response time when uploading and downloading data from Madrid to data containers deployed on Amazon AWS using  DynoStore (DS) with different storage and  Amazon S3.}
    \label{fig:diffs}
\end{figure}

Cloud providers like AWS offer access to various storage solutions. In this experiment, we evaluated DynoStore's performance in managing data containers using five different AWS storage options: \textit{i)} EBS-HDDs, \textit{ii)} EBS-SSD, \textit{iii)} FSx for Lustre, and \textit{iv)} a combination of all four configurations. These scenarios were tested under DynoStore's resilience configuration. For each scenario, we deployed up to 10 data containers on EC2 virtual machines (see Section~\ref{sec:infraestructure}). Additionally, we conducted the same experiments using Amazon S3 as a baseline for comparison.


Figure \ref{fig:diffs} shows on the y-axis the response time observed to upload and download different workload sizes (x-axis) under these scenarios. We uploaded and downloaded 100 files for each file size to obtain the average response time to service each request. We observe that for data sizes smaller than 1 GB, the response time for the SDD, HDD, and Lustre configurations is similar. This is because for small sizes, the data transfer time is short, and thus, the advantages (higher bandwidth and lower seek time) of SSD are not noticeable. For data sizes bigger than 1 GB, the benefits of having a parallel filesystem and higher I/O throughput are evident on the Lustre and SSD configurations. Furthermore, we can also observe that DynoStore, using a heterogeneous distributed storage, performs better than Amazon-S3, yielding a performance gain of 10\% when uploading 10 GB of data. This is important because it shows that we can achieve and even improve the time yielded by public cloud storage solutions by having distributed and heterogeneous storage systems that use the available resources across different facilities.

\subsection{Dynamic selection of resilience parameters on nodes with different failure rates } \label{sec:moreres}

\begin{table}[t]
        \caption{Percentage of data retained depending on the number of node failures.}
    \label{tab:node_failures}
    \centering
    \begin{adjustbox}{width=\linewidth}
    \begin{tabular}{c|ccccccc}
        \toprule
        \multirow{2}{*}{\textbf{Algorithm}} & \multicolumn{7}{c}{\textbf{Number of Failures}} \\ 
        \cline{2-8} 
        & \addstackgap{\textbf{0}} & \textbf{1} & \textbf{2} & \textbf{3} & \textbf{4} & \textbf{5} & \textbf{6} \\
        \midrule
        DynoStore & $100\%$ & $100\%$ & $100\%$ & $100\%$ & $100\%$ & $100\%$ & $40\%$ \\
        HDFS & $100\%$ & $100\%$ & $100\%$ & $100\%$ & $100\%$ & $60\%$  & $0\%$\\
        GlusterFS & $100\%$ & $100\%$ & $100\%$ & $100\%$ & $82\%$ & $0\%$  & $0\%$\\
        DAOS & $100\%$ & $90\%$ & $93\%$ & $93\%$ & $82\%$ & $0\%$  & $0\%$\\
        \bottomrule
    \end{tabular}
    \end{adjustbox}
\end{table}

DynoStore's modular design is flexible enough to include new data management strategies to improve requirements such as efficiency and resilience. 
Here, we describe an experiment in which we evaluate a dynamic algorithm that determines, in real-time, how many data and parity chunks to create and where to place them to maximize the number of node failures the data can withstand. This algorithm considers geographically distributed data containers and heterogeneous environments, assuming each container is prone to failure.

We conducted simulated experiments using a video dataset~\cite{Corona_2021_WACV} with a reliability target ensuring that each data item has a maximum probability of loss of 0.1\% over a 1-year period. The evaluation was performed in a scenario with ten heterogeneous data containers, each exhibiting an annual failure rate between 1\% and 25\%. We measured the percentage of data that remained accessible as the number of failed data containers increased.
Additionally, we compared DynoStore's performance against other resilience systems using Reed-Solomon erasure coding with their default configuration: HDFS (6 data blocks, 3 parity blocks), GlusterFS (4 data blocks, 2 parity blocks), and DAOS (8 data blocks, 2 parity blocks). Table \ref{tab:node_failures} presents the results. DynoStore is the only system capable of retaining all data even when 5 out of 10 storage nodes fail, guaranteeing that each data item has a maximum 0.1\%  probability of loss. This demonstrates that DynoStore's resilience algorithm can be adjusted to accommodate data container scenarios with varying resilience, storage, and latency characteristics.

\subsection{Case study I: Processing medical data} 

\begin{figure}
    \centering
    \includegraphics[width=0.78\linewidth]{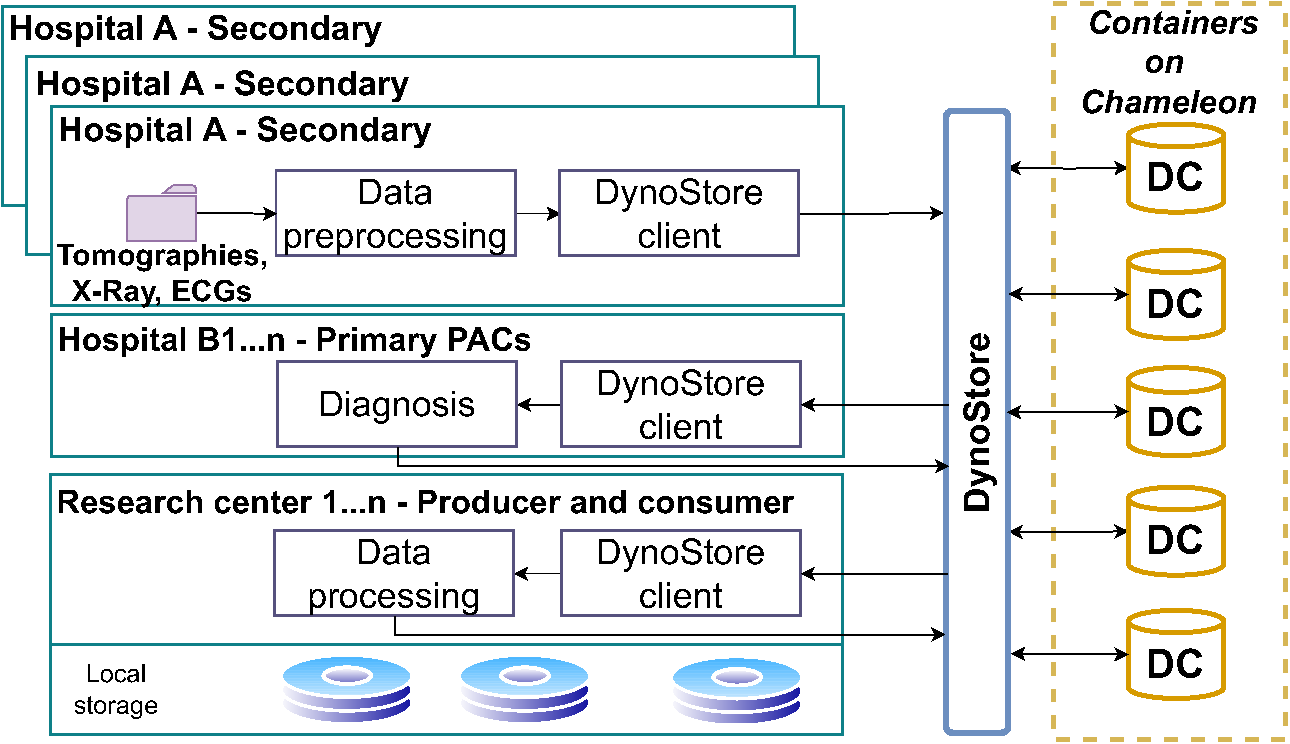}
    \caption{Design of medical data management case study.}
    \label{fig:usecase}
\end{figure}

Figure \ref{fig:usecase} shows the conceptual representation of a case study focused on managing medical data across different facilities to create a secure distribution network that helps physicians diagnose remote patients and research to help develop systems to help healthcare procedures. 

The application for this case study was developed using Globus Compute \cite{chard2020funcx} and ProxyStore \cite{pauloski2023accelerating}. Globus Compute is a FaaS platform that allows functions to be deployed and executed across distributed endpoints. ProxyStore manages data references for objects stored in DynoStore.

We evaluated the application's performance using the following data managers to move data through functions: DynoStore, Redis, and IPFS. Redis was configured to persist data by periodically backing up to disk and logging each operation. This setup ensured a fair comparison, as both systems were responsible for handling data transport and storage. In this case, Redis nodes are deployed in the same region of Chameleon, creating a cluster of virtual machines under the same network, which is a typical setup for this solution.

\begin{figure}
    \centering
    \includegraphics[width=0.75\linewidth]{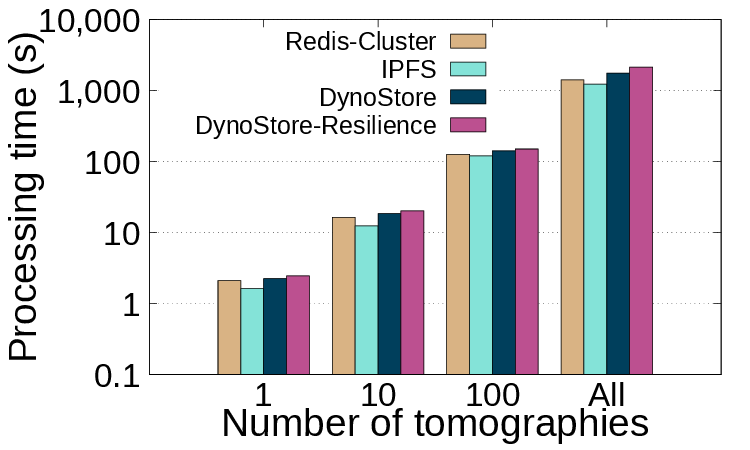}
    \caption{Response time observed when processing lung tomography images.}
    \label{fig:processingtomschame}
\end{figure}

Figure \ref{fig:processingtomschame} shows the total time (y-axis) for processing varying numbers of tomography images and all available images (x-axis). On average, each image has a size of 0.1 MB, and all datasets have a size of 2.1 GB. As can be observed,  IPFS is the solution that yields a lower processing time, as its P2P model does not rely on a centralized server to transfer data. In comparison, DynoStore and Redis are similar across all data sizes evaluated. For the full dataset (2.1 GB), IPFS spent 20.6 minutes, Redis took 23.5 minutes, DynoStore 29.4 minutes, and DynoStore with resilience configuration took 35.7 minutes. Here are two important aspects to consider. The first one is that IPFS is a P2P protocol that directly transfers data between two peers, and it does not implement an active replication of data for fault tolerance. Redis and DynoStore implement fault-tolerance strategies based on replication and erasure codes, respectively. The second aspect is that while DynoStore stores data across multiple and heterogeneous locations, Redis stores data in a local cluster, which reduces the overhead generated by data management tasks.



\subsection{Case study II: Management of satellite imagery through multiple locations}



Using DynoStore, ProxyStore, and Globus Compute, we designed an Earth observation system to process, store, and share data across multiple locations. This demonstrates DynoStore's flexibility in handling diverse scenarios and workloads, managing data across storage silos in different locations. 

\begin{figure}
    \centering
    \includegraphics[width=0.75\linewidth]{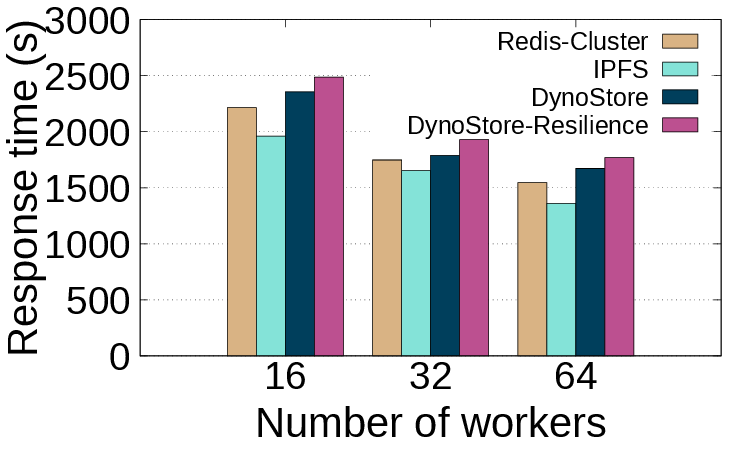}
    \caption{Response time observed for processing satellite images using a different number of workers.}
    \label{fig:parallel-satellite}
    \vspace{-1.5em}
\end{figure}

Figure \ref{fig:parallel-satellite} shows the response time (y-axis) for managing satellite images with different configurations and numbers of workers (x-axis). DynoStore again delivers competitive performance compared to Redis and IPFS for data transport. As expected, increasing the number of parallel workers deployed with Globus Compute reduces the response time. For example, comparing 16 workers with 64, there was observed a reduction in the response time of 28\%-30\% in all configurations. This parallel setup significantly improves system performance, allowing efficient management of large data volumes.


\section{Discussion} \label{sec:disc}



Here, we discuss the lessons learned from the evaluation conducted using DynoStore and state-of-the-art approaches.

\textbf{DynoStore and state-of-the-art approaches:} Through our benchmarks and case studies, we compared DynoStore's performance with state-of-the-art solutions like Amazon S3, Redis, and IPFS. Our results show that DynoStore performs competitive while automatically managing data in a heterogeneous environment. Key differences include flexibility and data management. Redis focuses on low-latency access from a single location but is not recommended for multi-location setups due to its reliance on a stable and low-latency network as well as the need for users to open ports for communication. Although Amazon S3 supports multiple endpoints, its integration with on-premises storage is complex, requiring gateways and clients. IPFS, with its decentralized P2P model, connects endpoints for data sharing but lacks management features like load balancing for efficient node capacity use.  While in some experiments, DynoStore is slower (i.e., compared with IPFS), our solution meets different needs in terms of reliability, resilience, and distribution. 

\textbf{Data resiliency:} In our evaluation, we found that DynoStore's resiliency policy incurs an overhead of about 11\% when uploading 100 GB of data. This overhead is offset by DynoStore's ability to maintain data access even when storage locations are unavailable. In comparison to other techniques like HDFS, DynoStore demonstrates lower storage overhead, with HDFS requiring 300\% overhead to tolerate two failures, while DynoStore only requires 20\%. IPFS does not replicate files until requested, which risks data unavailability if the storing node fails. To ensure redundancy, users must implement IPFS Cluster \cite{ipfs-cluster}, adding complexity through a new management layer. 

\textbf{Applicability to manage data in FaaS applications:} In two case studies---one for managing medical data and the other for satellite imagery---we demonstrated the development of two FaaS applications that process data across multiple sites combining DynoStore with ProxyStore and Globus Compute. 
The experimental evaluation showed that Redis outperforms DynoStore by 35\% compared to the DynoStore resilience configuration. However, Redis does not support data storage across multiple sites, as it relies on all nodes sharing a network for low-latency communication.

\section{Conclusions and future work}
\label{sec:conc}

DynoStore simplifies the integration of multiple storage systems into a unified solution using abstraction units called data containers. DynoStore leverages data containers for seamless integration, erasure coding for efficient fault tolerance, and Paxos-based consistency mechanisms to maintain strong guarantees in distributed environments. We implemented a DynoStore prototype and evaluated it under different data management scenarios. The flexibility of DynoStore in connecting different storage systems and managing data across them was demonstrated on different micro-benchmarks and applications. We have also shown in simulations that DynoStore's flexibility allows it to withstand more node failures than traditional filesystems. We showed that DynoStore can reliably store data with performance comparable to existing data management solutions.

While DynoStore provides strong consistency and fault tolerance for metadata through Paxos-based replication, future work will focus on integrating transactional metadata access and concurrency control mechanisms. Moreover, we are also investigating the adoption of advanced load-balancing techniques that consider bandwidth, energy, and budget constraints. Furthermore, we are exploring hybrid resilience strategies combining replication and erasure codes to further improve performance and adaptability.  We will investigate techniques for making intelligent decisions about moving data near computation or vice versa. DynoStore advances the state of the art in distributed storage systems by combining scalable data containers, efficient load-balancing algorithms, and resilient erasure coding mechanisms, offering the integration of across heterogeneous storage systems while efficiently using available resources. 



\bibliographystyle{IEEEtran}
\bibliography{references}

\begin{thebibliography}{10}
\providecommand{\url}[1]{#1}
\csname url@samestyle\endcsname
\providecommand{\newblock}{\relax}
\providecommand{\bibinfo}[2]{#2}
\providecommand{\BIBentrySTDinterwordspacing}{\spaceskip=0pt\relax}
\providecommand{\BIBentryALTinterwordstretchfactor}{4}
\providecommand{\BIBentryALTinterwordspacing}{\spaceskip=\fontdimen2\font plus
\BIBentryALTinterwordstretchfactor\fontdimen3\font minus
  \fontdimen4\font\relax}
\providecommand{\BIBforeignlanguage}[2]{{%
\expandafter\ifx\csname l@#1\endcsname\relax
\typeout{** WARNING: IEEEtran.bst: No hyphenation pattern has been}%
\typeout{** loaded for the language `#1'. Using the pattern for}%
\typeout{** the default language instead.}%
\else
\language=\csname l@#1\endcsname
\fi
#2}}
\providecommand{\BIBdecl}{\relax}
\BIBdecl

\bibitem{boubiche2018big}
S.~Boubiche, D.~E. Boubiche, A.~Bilami, and H.~Toral-Cruz, ``Big data
  challenges and data aggregation strategies in wireless sensor networks,''
  \emph{IEEE access}, vol.~6, pp. 20\,558--20\,571, 2018.

\bibitem{stephens2015big}
Z.~D. Stephens, S.~Y. Lee, F.~Faghri \emph{et~al.}, ``Big data: astronomical or
  genomical?'' \emph{PLoS biology}, vol.~13, no.~7, p. e1002195, 2015.

\bibitem{dash2019big}
S.~Dash, S.~K. Shakyawar, M.~Sharma, and S.~Kaushik, ``Big data in healthcare:
  management, analysis and future prospects,'' \emph{Journal of big data},
  vol.~6, no.~1, pp. 1--25, 2019.

\bibitem{ahmadian2018investigating}
S.~Ahmadian, F.~Taheri, M.~Lotfi, M.~Karimi, and H.~Asadi, ``Investigating
  power outage effects on reliability of solid-state drives,'' in \emph{2018
  DATE}.\hskip 1em plus 0.5em minus 0.4em\relax IEEE, 2018, pp. 207--212.

\bibitem{draganski2019nation}
B.~Draganski, F.~Kherif, D.~Damian, J.-F. Demonet \emph{et~al.}, ``A
  nation-wide initiative for brain imaging and clinical phenotype data
  federation in swiss university memory centres,'' \emph{Current opinion in
  neurology}, vol.~32, no.~4, pp. 557--563, 2019.

\bibitem{kahn2011future}
S.~D. Kahn, ``On the future of genomic data,'' \emph{science}, vol. 331, no.
  6018, pp. 728--729, 2011.

\bibitem{kotsev2020spatial}
A.~Kotsev, M.~Minghini, R.~Tomas \emph{et~al.}, ``From spatial data
  infrastructures to data spaces—a technological perspective on the evolution
  of european sdis,'' \emph{ISPRS International Journal of Geo-Information},
  vol.~9, no.~3, p. 176, 2020.

\bibitem{mishra2023advanced}
P.~Mishra, ``Advanced aws services,'' in \emph{Cloud Computing with AWS:
  Everything You Need to Know to be an AWS Cloud Practitioner}.\hskip 1em plus
  0.5em minus 0.4em\relax Springer, 2023, pp. 247--277.

\bibitem{opara2016critical}
J.~Opara-Martins, R.~Sahandi, and F.~Tian, ``Critical analysis of vendor
  lock-in and its impact on cloud computing migration: a business
  perspective,'' \emph{Journal of Cloud Computing}, vol.~5, pp. 1--18, 2016.

\bibitem{liu2017minimum}
G.~Liu and H.~Shen, ``Minimum-cost cloud storage service across multiple cloud
  providers,'' \emph{IEEE/ACM Transactions on Networking}, vol.~25, no.~4, pp.
  2498--2513, 2017.

\bibitem{suter2023driving}
F.~Suter, R.~F. Da~Silva, A.~Gainaru, and S.~Klasky, ``Driving next-generation
  workflows from the data plane,'' in \emph{19th e-Science}.\hskip 1em plus
  0.5em minus 0.4em\relax IEEE, 2023, pp. 1--10.

\bibitem{aghayev2019file}
A.~Aghayev, S.~Weil, M.~Kuchnik \emph{et~al.}, ``File systems unfit as
  distributed storage backends: lessons from 10 years of ceph evolution,'' in
  \emph{27th ACM SOSP}, 2019, pp. 353--369.

\bibitem{madireddy2018machine}
S.~Madireddy, P.~Balaprakash, P.~Carns \emph{et~al.}, ``Machine learning based
  parallel i/o predictive modeling: A case study on lustre file systems,'' in
  \emph{ISC High Performance 2018}.\hskip 1em plus 0.5em minus 0.4em\relax
  Springer, 2018, pp. 184--204.

\bibitem{ghazi2015hadoop}
M.~R. Ghazi and D.~Gangodkar, ``Hadoop, mapreduce and hdfs: a developers
  perspective,'' \emph{Procedia Computer Science}, 2015.

\bibitem{bornholt2021using}
J.~Bornholt, R.~Joshi, V.~Astrauskas \emph{et~al.}, ``Using lightweight formal
  methods to validate a key-value storage node in amazon s3,'' in \emph{ACM
  SIGOPS}, 2021.

\bibitem{zhang2020exploring}
Z.~Zhang, B.~Bockelman, D.~Weitzel, and D.~Swanson, ``Exploring erasure coding
  techniques for high availability of intermediate data,'' in \emph{20th
  IEEE/ACM CCGrid}.\hskip 1em plus 0.5em minus 0.4em\relax IEEE, 2020, pp.
  865--872.

\bibitem{garcia2003design}
F.~Garcia, A.~Calderon, J.~Carretero \emph{et~al.}, ``The design of the expand
  parallel file system,'' \emph{The International Journal of High Performance
  Computing Applications}, vol.~17, no.~1, pp. 21--37, 2003.

\bibitem{munoz2024fault}
Mu{\~n}oz-Mu{\~n}oz, Garcia-Carballeira, Camarmas-Alonso \emph{et~al.}, ``Fault
  tolerant in the expand ad-hoc parallel file system,'' in
  \emph{Euro-PAR}.\hskip 1em plus 0.5em minus 0.4em\relax Springer, 2024, pp.
  62--76.

\bibitem{hennecke2020daos}
M.~Hennecke, ``Daos: A scale-out high performance storage stack for storage
  class memory,'' \emph{Supercomputing frontiers}, vol.~40, 2020.

\bibitem{cloudfront}
``Amazon cloudfront,'' \url{https://aws.amazon.com/es/cloudfront/}, {Accessed:
  31/05/2024}.

\bibitem{azure}
``Azure content delivery network,''
  \url{https://azure.microsoft.com/es-es/products/cdn}, {Accessed: 31/05/2024}.

\bibitem{fastly}
``Fastly,'' \url{https://www.fastly.com/products/cdn}, {Accessed: 31/05/2024}.

\bibitem{benet2014ipfs}
J.~Benet, ``Ipfs-content addressed, versioned, p2p file system,'' \emph{arXiv
  preprint arXiv:1407.3561}, 2014.

\bibitem{barisits2019rucio}
M.~Barisits, T.~Beermann, F.~Berghaus \emph{et~al.}, ``Rucio: Scientific data
  management,'' \emph{Computing and Software for Big Science}, vol.~3, pp.
  1--19, 2019.

\bibitem{irods}
M.~Hedges, A.~Hasan, and T.~Blanke, ``Management and preservation of research
  data with irods,'' in \emph{ACM CIMS}.\hskip 1em plus 0.5em minus 0.4em\relax
  New York, NY, USA: Association for Computing Machinery, 2007, p. 17–22.

\bibitem{kubiatowicz2000oceanstore}
J.~Kubiatowicz, D.~Bindel, Y.~Chen, S.~Czerwinski \emph{et~al.},
  ``Ocean{S}tore: An architecture for global-scale persistent storage,''
  \emph{ACM SIGOPS Operating Systems Review}, vol.~34, no.~5, pp. 190--201,
  2000.

\bibitem{howard2020paxos}
H.~Howard and R.~Mortier, ``Paxos vs raft: Have we reached consensus on
  distributed consensus?'' in \emph{7th PaPoC}, 2020, pp. 1--9.

\bibitem{carrizales2024structmesh}
D.~Carrizales-Espinoza, D.~D. Sanchez-Gallegos, J.~Gonzalez-Compean, and
  J.~Carretero, ``Structmesh: A storage framework for serverless computing
  continuum,'' \emph{FGCS}, 2024.

\bibitem{rabin1990information}
M.~O. Rabin, ``The information dispersal algorithm and its applications,'' in
  \emph{Sequences: Combinatorics, Compression, Security, and
  Transmission}.\hskip 1em plus 0.5em minus 0.4em\relax Springer, 1990, pp.
  406--419.

\bibitem{chard2017globus}
K.~Chard, I.~Foster, and S.~Tuecke, ``Globus: Research data management as
  service and platform,'' in \emph{Proceedings of the Practice and Experience
  in Advanced Research Computing 2017 on Sustainability, Success and Impact},
  2017, pp. 1--5.

\bibitem{pauloski2023accelerating}
J.~G. Pauloski, V.~Hayot-Sasson, L.~Ward \emph{et~al.}, ``Accelerating
  communications in federated applications with transparent object proxies,''
  in \emph{Supercomputing}, 2023.

\bibitem{lctsc}
J.~Yang, G.~Sharp, H.~Veeraraghavan \emph{et~al.}, ``{Data from Lung CT
  Segmentation Challenge (LCTSC) (Version 3) [Data set].}'' \emph{The Cancer
  Imaging Archive}, 2017.

\bibitem{li2016data}
Li, Abramson, Arlinghaus \emph{et~al.}, ``{Data from QIN-breast},'' \emph{The
  Cancer Imaging Archive}, 2016.

\bibitem{keahey2020lessons}
K.~Keahey, J.~Anderson, Z.~Zhen \emph{et~al.}, ``Lessons learned from the
  chameleon testbed,'' in \emph{USENIX ATC '20}.\hskip 1em plus 0.5em minus
  0.4em\relax USENIX Association, July 2020.

\bibitem{Corona_2021_WACV}
K.~Corona, K.~Osterdahl, R.~Collins, and A.~Hoogs, ``Meva: A large-scale
  multiview, multimodal video dataset for activity detection,'' in
  \emph{IEEE/CVF WACV}, January 2021, pp. 1060--1068.

\bibitem{chard2020funcx}
R.~Chard, Y.~Babuji, Z.~Li, T.~Skluzacek, A.~Woodard, B.~Blaiszik, I.~Foster,
  and K.~Chard, ``Funcx: A federated function serving fabric for science,'' in
  \emph{29th HPDC}, 2020, pp. 65--76.

\bibitem{ipfs-cluster}
IPFS, ``Ipfs cluster,'' \url{https://github.com/ipfs-cluster/ipfs-cluster},
  2024.

\end{thebibliography}

\end{document}